# Effect of grain disorientation on early fatigue crack propagation in FCC poly-crystals: a three dimensional dislocation dynamics investigation

G.V. Prasad Reddy[a,b], C. Robertson[c], C. Déprés[a], M. Fivel[d]

[a]*Laboratoire SYstèmes et Matériaux pour la MEcatronique, Université de Savoie, BP80439, 74944 Annecy-le-Vieux Cedex, France*

[b]*Mechanical Metallurgy Division, Indira Gandhi Center for Atomic Research, Kalpakkam, Tamil Nadu 603102 INDIA*

[c]*CEA, DEN, Service de Recherche Métallurgiques Appliquées, F91191 Gif-sur-Yvette, France*

[d]*SIMaP-GPM2, Grenoble INP, CNRS/UJF, 101 Rue de la Physique, BP 46, 38402 St Martin d'Hères cedex, France*

Keywords: Discrete dislocation dynamics; Fatigue; Crack propagation; Crack-tip stress, Grain boundaries

## Abstract

Three-dimensional dislocation dynamics simulations are used to study micro-crack interaction with the first micro-structural barrier in face-centred cubic bi-crystals loaded in high cycle fatigue conditions. In the examined configuration, we assumed that micro-crack transmission occurs due to surface relief growth in the secondary grain ahead of the primary crack. This indirect transmission mechanism is shown to strongly depend on grain-1/grain-2 disorientation. For instance, small grain disorientation induces plastic strain localisation ahead of the crack and faster transmission through the first barrier. Conversely, large grain-1/grain-2 disorientation induces plastic strain spreading similar to crack tip blunting yielding slower indirect transmission. A semi-analytical micro-model is developed based on the present simulation results and complementary experimental observations highlighting the original notion of first-barrier compliance. The model captures well known experimental trends including effects of: grain-size, grain disorientation and micro-crack retardation at the first barrier.

## 1. Introduction

Initiation of stage I fatigue cracks takes place in surface grains of single-phased, face-centred cubic (FCC) poly-crystals (in absence of defects and/or precipitates). Micro-cracks initiation involves surface marking development in the form of extrusions, associated with persistent slip bands (PSB) [1-3]. Micro-cracks initiate after sufficient extrusion growth, corresponding to the accumulation of a critical amount of plastic slip $\gamma_{lim}$. The later quantity is a material parameter, depending on the environment and test temperature conditions [4-6]. Micro-cracks then grow up to the first micro-structural barrier *i.e.* the grain boundary. At this stage, subsequent propagation can be retarded during a certain time, depending on the loading amplitude and the orientation of the next (or secondary) grain [7]. In any case, plasticity developing ahead of the first micro-structural barrier is a crucial and yet poorly understood phenomenon, controlling the high cycle fatigue lifetime (HCF) [8-12].

Experimental observations can help finding the micro-crack transmission mechanism that is best adapted to the later configuration. In practice, fracture surfaces of stage I cracks are semi-elliptic in shape, i.e. elongated parallel to the specimen surface [13-15]. This means transmission of a primary micro-crack is always faster towards a neighbouring surface grain than towards a neighbouring bulk grain [16,17]. This effect can be ascribed to the larger crack tip displacement amplitude observed in the surface region, where the crystal is less constrained than in the bulk [18-22]. It is sometimes assumed that the crack front and the associated slip is directly transmitted to the next surface grain [11,16], as sketched in Fig.1a. The particular slip direction **b** of active slip systems in primary grains is, however, detrimental to direct slip transmission [20]. Indeed, the slip direction **b** in the primary grains is roughly perpendicular to the free surface, yielding a fast extrusion growth. Correspondingly, this means **b** is more or less parallel to the first grain boundary, which is very unfavourable for direct slip transmission and subsequent crack front propagation. This slip transmission is even less probable in HCF, where the applied loading amplitude is limited.

Our goal in this work is to investigate plasticity mechanisms affecting early micro-crack propagation, in HCF conditions. The implemented simulation setup is adapted to study a particular three-dimensional mechanism to be called *indirect transmission* as defined in Fig.1b. This mechanism specifically applies to micro-crack interaction with the first micro-structural barrier, where the initiated cracks can spend a large part of their growth time. The adopted DD simulation setup is based on the various, above-described experimental observations. For this and other reasons, our simulations do not apply to prior or subsequent micro-crack development stages and substantially differ from the previous crack-tip plasticity investigations [24-27].

**2 DD model setup for indirect transmission investigation**

*2.1 Boundary conditions, simulation cell*

The 3D dislocation dynamics code used for the study is the edge-screw model called TRIDIS, developed at SIMaP laboratory [28]. Plastic strain is carried by dislocation lines discretized into edge and screw segments which glide on a discrete lattice, homothetic to the underlying crystallographic structure. From the mechanical point of view, the dislocation lines are treated as line singularities embedded in an elastic medium. Each dislocation segment generates a long range stress field in the entire simulated volume. Materials parameters of austenitic stainless steel are used herein, as listed in Table 1. Thermally-activated cross-slip is implemented as explained in [29,30].

A pseudo bi-grain aggregate simulation setup is adopted, with a view to investigate the indirect crack transmission mechanism as presented in Section 1 and in Fig.1b. The primary grain, also denoted as *grain-1*, includes an arrested crack (see also Section 2.2 below) and contains no dislocations. Note that grain-1 geometry is arbitrary and has no effect on this study. The hexagonal shape represented in Fig.1 is adopted for simplicity. The DD simulated space corresponds to the secondary grain, also denoted as *grain-2*, defined as a 5 wedged-cylinder. The cylinder-grain axis is aligned perpendicularly to the gain free surface (see Fig.1). Dislocations gliding in grain-2 are free to escape through the top surface of the simulated space, whereas all the other surfaces act as strong, impenetrable obstacles for dislocations. Each dislocation segment leaving the crystal through the top boundary prints a step in the corresponding surface. Each dislocation-induced surface step has a height of b, the magnitude of the Burgers vector. This effect of plasticity on surface topography is computed here using the same post-treatment method as described in [31]. In the present simulations, image forces are not implemented, since their influence on PSB slip activity has been proved to be negligible [31,32], including in presence of a crack [33].

Elastic and plastic incompatibilities between the investigated bi-grain aggregate and its environment, *i.e.* the un-cracked neighbouring grains, are not accounted for. The authors believe that this is a reasonable assumption as per the following arguments: i- Fatigue simulations carried out in single surface grains yield exactly the same dislocation microstructures and stress-strain response as observed in actual fatigued poly-crystals [29]. Neighbouring grain incompatibilities thus appear as a second order effect on the plasticity mechanisms associated with HCF conditions. ii- Stress modulations due to neighbouring grain elastic incompatibilities are much smaller (20% maximum, in anisotropic elasticity framework [34]) than those due to the primary crack stress field.

Plastic incompatibilities between cracked grain-1 and un-cracked grain-2 generates confined slip activity ahead of the crack tip, i.e. within grain-2 [34-38]. In actual poly-crystals, dislocation sources in the secondary grain can be directly punched through or nucleated at grain-1/grain-2 boundary. Here, we do not distinguish between either mechanism: the initial dislocation sources are simply placed within 1 μm of the arrested crack tip (in grain-2) and evolve according to the local, effective stress conditions (including the primary crack stress field).

The applied stress level is varied stepwise, throughout the simulated fatigue cycles. Whenever a load step is enforced, plastic strain then develops until the whole dislocation microstructure is in equilibrium with the external loading, as explained in [29]. Numerically, quasi-static loading conditions are achieved by keeping the load constant while performing discrete *time* steps until the resulting plastic strain *rate* becomes lower than a pre-selected value. The adopted simulation setup (source positioning and quasi-static loading) yields realistic stress-strain behaviour, including in presence of geometrically singular loading conditions [33,39].

The initial, grain-2 dislocation microstructure includes 24 Frank-Read sources, 2 sources on each of the 12 FCC slip systems a/2<110>{111}. The selected number of sources is sufficient to generate homogeneous crack plastic zone throughout the simulated time. Adding more sources has no effect on subsequent plasticity development, since most of the sources are generated afterwards through cross-slip mechanism [29]. The fatigue simulations are carried out with a fully reversed cyclic deformation for an applied plastic strain range of $\Delta\varepsilon_p=10^{-4}$ and $2\times10^{-4}$ compatible with HCF conditions. All the simulations are performed under isothermal conditions at 300K.

*2.2 Heterogeneous crack stress field implementation*

The primary crack included in grain-1 generates a long range stress field in the surrounding elastic medium, comprising grain-2. This effect is treated using the superimposition principle where the total stress applied to each dislocation segment (within grain-2) is the sum of: (i) the homogeneous applied stress, plus, (ii) the dislocation-induced internal stress, plus, (iii) the crack-induced heterogeneous stress, calculated using Eqs.1-4 below.

The primary crack habit plane is favourably oriented for shear stress, i.e. at 45° with respect to the free surface and loading direction. In the present case, grain-1 crack stress field is a combination of mode I and mode II loading components [18]. For mode I loading component, we use the following 2D expression [40]:

$$\begin{bmatrix} \sigma_{xx} \\ \sigma_{yy} \\ \tau_{xy} \end{bmatrix} = K_I \frac{\cos \theta/2}{\sqrt{2\pi r}} \begin{bmatrix} 1 - \sin \theta/2 \sin 3\theta/2 \\ 1 + \sin \theta/2 \sin 3\theta/2 \\ \sin \theta/2 \sin 3\theta/2 \end{bmatrix} \quad (1)$$

where $\theta$ is the azimuth angle (in grain-2: -90° < $\theta$ < +90°), r is the distance to crack tip and

$$K_I = \sigma_{xx(r \to \infty)} \sqrt{\pi a} \quad (2)$$

where a stands for the crack length and the reference coordinate system (x,y) is set normal to the crack front. A similar formulation is used to treat mode II loading component [40]:

$$\begin{bmatrix} \sigma_{xx} \\ \sigma_{yy} \\ \tau_{xy} \end{bmatrix} = K_{II} \frac{1}{\sqrt{2\pi r}} \begin{bmatrix} -\sin \theta/2 \, (2 + \cos \theta/2 \cos 3\theta/2) \\ \cos \theta/2 \sin \theta/2 \cos 3\theta/2 \\ \cos \theta/2 \, (1 - \sin \theta/2 \sin 3\theta/2) \end{bmatrix} \quad (3)$$

with:

$$K_{II} = \sigma_{yy(r \to \infty)} \sqrt{\pi a} \quad (4)$$

Eqs. 1-4 are calculated in consistence with grain-1 crack position and orientation, i.e. position r = 0 coincides with grain-1/grain-2 boundary (r < 0 fall within grain-1, r > 0 fall within grain-2). For analysis purpose, different DD simulations are performed for various orientations of grain-2, with respect to the primary crack plane. Each orientation is characterized by tilt and twist angles sketched in Fig.2 as defined in [16]. The simulations results are presented qualitatively, in Section 3 and further analyzed quantitatively, in Section 4.

## 3. Results

*3.1 Effect of heterogeneous crack stress field on slip system selection effect in grain-2*

Typical simulation results are presented below for 2 typical orientation of the secondary grain. In simulation 1 (see Fig.3), the grain orientation is characterized by tilt = -18° and twist = 0°. Slip activity is evidenced in only two slip systems denoted as B4 and D4 (following the usual Schmid and Boas nomenclature. Slip system B4 is considered as dominant, because dislocation density $\rho_{B4} > \rho_{D4}$ throughout the simulated time. The observed active slip systems are consistent with Fig.3b and 3c data showing the resolved shear stresses acting in each slip system of grain-2 in presence of the grain-1 crack. In Fig.3b, the maximal loading amplitude is found in system C3 (for azimuth > +70°). However; systems B4 and D4 are activated instead of C3. This effect directly relates to plane B and D crystallographic orientation. Indeed, Fig. 3c shows that planes B and D have a minimal twist angle, with respect to the arrested crack plane. Moreover, the corresponding plane B and D azimuth angles fall wherever the resolved shear stress is strong, yielding maximum slip activity. On the other hand, plane C exhibits a large twist angle, so dislocations moving on these planes rapidly roll away from the maximum stress azimuth.

In simulation 2 (Fig.4), the secondary grain orientation is characterized by tilt = 0°, twist = -35°. It leads to slip activity in systems A6 and D6 plus a slight activity in system C1, as shown in Fig.4a. Plane C is significantly more twisted with respect to the primary crack plane than both A and D planes (Fig.4c and 4d). Dislocations moving into planes A, D or C come across different azimuth angles and therefore, different resolved shear stresses. The dominant slip plane (D) corresponds to the least twisted of the 3 active ones, so that the resolved shear stress remains high, regardless of the azimuth angle.

These two examples of DD simulations reveal that the crack-induced stress field activates slip systems in grain-2 with any tilt and twist angle combination. In contrast, it has been shown that under homogeneous stress field conditions, the slip systems having a twist angle exceeding 70° are never activated [41]. Slip system activity thereby strongly depends on grain-1 crack stress field. This means that slip system selection may not be in accordance with continuum mechanics predictions based on Schmid law or stereographic projections.

*3.2 Effect of initial dislocation structure (due to pre-cycling)*

The effect of grain-1 crack stress field is examined in this section. In particular, we investigate whether grain-1 crack concentrates cyclic slip in existing grain-2 slip bands or promotes the formation of new slip bands, in entirely different (and initially inactive) slip systems. Hence, 2 distinct simulation setups are compared and analyzed in this section. In the "no-crack" setup ❶, a PSB dislocation structure is first generated in grain-2, by pre-deforming the grain during several cycles (typically 20-30 cycles) in absence of the cracked grain-1 (Fig.5c). Then, grain-2 (with a prior substructure) is cyclically loaded under the influence of grain-1 crack stress field. This will be referred as "crack and substructure" simulation ❷ as depicted in 5d.

In the "crack" simulation setup ❸ (Fig.5a and 5e), we use the same crack stress field as in simulation ❷ although this time, the cracked grain-1 is present right from the start of the simulation. The (somewhat unrealistic) simulation ❸ setup helps in highlighting the differences between "crack" and "crack and substructure" simulations: i.e., to check whether the cyclic slip remains concentrated in pre-existing PSB substructures or develop into different ones.

Actually, both situations are observed during our simulations, depending on the presence/absence of a common slip system between "with-crack" (WC) and "without-crack" (WOC) simulations. If the crack stress field (WC simulation ❸) activates exactly the same slip system as found in a WOC simulation ❶, then cyclic slip in simulation ❷ further concentrates within the prior substructure, leading to accelerated surface relief growth in the common slip system (not shown). When the active slip system differs between WOC and WC simulations, the prior dislocation substructure has little influence on subsequent surface relief evolutions (compare Fig.5d and 5e). In other words, grain-1 crack stress field can selectively activate slip systems in grain-2 regardless of the pre-existing dislocation substructures. In all the cases, there is a marked spatial distribution of plastic displacement irreversibility along the 3D crack front (not shown). For instance, slip irreversibility is more pronounced in grain-2 surface (curves ❶ or ❸ in Fig.5a) than anywhere else in the grain [33]. Interestingly, such slip repartition is consistent with the indirect crack transmission mechanism observed during the passage of the first-barrier (see Section 1). Finally, the degree of plastic strain localization in grain-2 directly depends on grain-1/grain-2 disorientation. This effect is examined in details, in the next section.

*3.3 Effect of grain-1/grain-2 disorientation: tilt and twist angles*

In absence of a crack in grain-1, dislocation structures develop in the entire grain-2, in the form of regularly spaced persistent slip bands [29,31]. In presence of a crack in grain-1 however, grain-2 orientation determines whether plasticity is localized or homogeneous ahead of the crack. In Figs. 6a and 6b for example, plastic strain is localised into a few pronounced slip bands, generating sharp and fast growing surface relief. In Figs.6c and 6d in contrast, plastic strain spreads more homogeneously, generating comparatively smoother and slower relief accumulation. This means that indirect micro-crack transmission (see Section 1) should be faster in Figs. 6a and 6b than in Figs. 6c and 6d. In other words, Fig.6 qualitatively shows that low-tilt and low-twist conditions accelerate indirect transmission; whenever the corresponding plastic deformation is localized (in grain-2).

These results will be further analyzed using a semi-quantitative model presented in next section.

**4. Discussion: micro-crack transmission model based on DD results**

When the applied stress is increased, the mobile dislocations present in the crack tip region of grain-2 glide into many different, parallel slip planes and gradually pile-up at the grain boundaries. Note that this configuration is observed in both our simulations (see Figs. 5 and 6) and in actual fatigued grains [37,38,42-44]. The stress-strain response associated with the crack tip region dislocation structures can be calculated by considering $\mathcal{N}$ slip planes[1], containing mobile dislocations in equilibrium with: i- the applied stress, ii- the line tension self-stress and iii- the mutual slip band interactions. That configuration can be described by a set of $\mathcal{N}$ equations, for which the solution is:

$$\Delta\tau_{local,wc} = \frac{1}{\mathcal{N}}\left[\frac{1}{S}\frac{\mu}{(1-\nu)}\right]\Delta\gamma_{local,wc} \quad (5)$$

where $\Delta\tau_{local,wc}$ and $\Delta\gamma_{local,wc}$ are the shear stress and shear strain ranges acting in the crack tip region (subscript reads: "local, with crack"), S a dimensionless factor characterizing the grain geometry, $\mu$ the shear modulus and $\nu$ the Poisson ratio. All the details of expression (5) derivation can be found in [32,45]. In the next two paragraphs, the different terms in Eq.5 will be evaluated separately, in an attempt to develop a semi-analytical, micro-crack transmission model.

*Evaluation of Eq.5: the left-hand term.*

The crack tip shear stress range is calculated using crack tip stress field expressions Eqs.1-4:

$$\Delta\tau_{local,wc} \approx \frac{\Delta K_0}{\sqrt{2\pi r}} = \Delta\tau^*_{global,wc}\sqrt{\frac{\pi a}{2\pi r}} \quad (6)$$

where $\Delta\tau^*_{global,wc}$ is the shear stress cumulated beyond the yield point (see Fig.5b), at the grain scale (and hence the subscript "global"). Setting the crack size a ~ $D_g$, i.e. assuming grain-1 is fully cracked, one finally obtains:

$$\Delta\tau_{local,wc} \approx \Delta\tau^*_{global,wc}\sqrt{\frac{D_g}{2r}} \quad (7)$$

*Evaluation of Eq.5: the right-hand term.*

---

[1] The number of active slip planes $\mathcal{N}$ must not be confused with the number of cycles $N$.

In Section 3, it is shown that the presence of a primary crack strongly accelerates the surface relief development within grain-2 (see ❶→❷ transition, in Fig.5a). This effect can be ascribed to the particular stress-strain conditions acting in the (near surface) crack tip region. Namely, the applied fatigue cycles generate asymmetric plastic displacements at this location (see Section 3.2) which correspond to positive mean strain conditions. In other words, the plastic strain range $\Delta\gamma_{local,wc}$ (as defined in Eq.5) corresponds to a given local mean strain level $\overline{\gamma}_p$. Quantity $\Delta\gamma_{local,wc}$ cannot be directly calculated using $\Delta\gamma_{global,wc}$ (see curve ❷ in Fig.5a), or $\Delta\varepsilon_p$ as shown in Fig.5b. Rather, $\Delta\gamma_{local,wc}$ needs to be computed through its consequences in terms of surface relief evolutions. To make this calculation we need first, to bring in the general expressions describing the surface displacement evolutions $\gamma^{surf}$ in absence and in presence of a mean plastic strain $\overline{\gamma}_p$.

In absence of a mean strain (tension = compression), cyclic surface displacement accumulation is described by [31]:

$$\gamma^{surf} \propto \eta \Delta\gamma_p \sqrt{N} \tag{8}$$

where $\eta$ is a dimensionless quantity depending on simulation setup and material parameters (see Table 1), $\Delta\gamma_p$ the imposed plastic (shear) strain range and $N$ the number of cycles (see reference [31] for derivation of Eq. 8 using fatigue DD simulations).

In presence of a mean strain (tension ≠ compression), cyclic surface displacement accumulation is given by:

$$\gamma^{surf} \propto \eta \Delta\varepsilon_p (1 + 2\frac{\overline{\gamma}_p}{\Delta\gamma_p})\sqrt{N} \tag{9}$$

Derivation of Eq.9 is based on a set of cyclic simulations, where the mean strain level $\overline{\gamma}_p$ is systematically varied. These results and their physical interpretation were presented in reference [46]. In practice, quantity $\overline{\gamma}_p$ due to the primary crack can then be evaluated using two distinct DD simulations, using a fixed plastic strain range $\Delta\gamma_p$. A first simulation, carried out without crack, generates a reference $\gamma_{woc}^{surf}$ curve (similar to curve ❶ in Fig.5a). This curve is then fitted using Eq.8 (as shown in Fig.8a of reference [29]), yielding a definite $\eta$ value. A second simulation, performed in presence of a primary crack, is used to generate a $\gamma_{wc}^{surf}$ curve (similar to curve ❸ in Fig.5a). That new curve is fitted[2] using Eq.9, i.e. assuming that the differences between $\gamma_{wc}^{surf}$ and $\gamma_{woc}^{surf}$ is due to a mean strain $\bar{\gamma}_p$ acting in the crack tip region. The resulting $\bar{\gamma}_p$ value is finally inserted in Eq.5, by setting $\Delta\gamma_{local,wc} = \bar{\gamma}_p$.

*Micro-mechanical model application to indirect micro-crack transmission.*

Inserting Eq.7 in the left-hand term of Eq.5 gives, after re-arranging the different terms:

$$\frac{D_g}{(2r)} = \left(\frac{1}{\mathcal{N}}\left[\frac{1}{S}\frac{\mu}{(1-\nu)}\right]\frac{\Delta\gamma_{local,wc}}{\Delta\tau^*_{global,wc}}\right)^2 \qquad (10)$$

Eq.10 description is tested hereafter through various DD simulations, carried out using different twist angles[3]: 20°, 38°, 71° and 82°, for a fixed cyclic loading level $\Delta\gamma_p = 10^{-4}$. Quantity $\Delta\tau^*_{global,wc}$ depends on grain-1/ grain-2 disorientation. Thus, one $\Delta\tau^*_{global,wc}$ value is determined per each test case using the corresponding cyclic stress-strain data (see Fig.5b for example). Surface displacement evolutions obtained in each case (see Fig.5a for example) also correspond to a definite $\Delta\gamma_{local,wc}$ value, calculated using Eqs.8-9. To each test-case thus corresponds: i- a different $\Delta\gamma_{local,wc}/\Delta\tau^*_{global,wc}$ ratio, which we call the *first-barrier compliance*, ii- a different curve in Fig.7. For a fixed $D_g/2r$ value, it is readily seen that number of active slip planes $\mathcal{N}$ in the crack tip zone increases with the twist angle (see Fig. 7). In other words, Eq.10 description is fully consistent with the simulation results presented in Section 3.3 (see Fig.6).

In actual poly-crystals, micro-crack transmission is a local process, taking place at the scale of individual shear bands [16]. Surface displacements due to individual shear bands thus need to be evaluated, in order to assess the kinetics of indirect micro-crack transmission. From Eq.8 and Eq.9, we know that surface displacement evolution in the whole crack tip region is characterized by:

$$\frac{\gamma_{wc}^{surf}}{\gamma_{woc}^{surf}} = (1 + 2\frac{\overline{\gamma_p}}{\Delta\gamma_p}) \tag{11}$$

---

[2] Taking $\eta$ from the reference, crack-less simulation.
[3] We used a fixed tilt angle: 24±2° in all the cases.

Surface displacement due to a single shear band can be calculated by dividing the right-hand term of Eq. 11 by the total number of active shear bands $\mathcal{N}$, which can be obtained by solving Eq.10. In this way, the following expression is readily obtained:

$$\frac{\gamma_{wc}^{surf}}{\gamma_{woc}^{surf}} = \frac{\sqrt{\frac{D_g}{2r}}}{\left[\frac{1}{S(1-\nu)}\right]\left(\frac{\Delta\eta_{local,wc}}{\Delta\tau_{global,wc}^*}\right)}\left(1 + \sqrt{\frac{D_g}{2r}}\right) \tag{12}$$

Eq.12 describes the surface displacement accumulation due to a single shear band located in grain-2 ahead of the primary crack tip. Indirect micro-crack transmission takes place when the surface displacement accumulation $\gamma_{wc}^{surf}$ (from Eq.12) becomes critically large (within grain-2). If we set $\gamma_{wc}^{surf} = \gamma_{lim}$ and take $\gamma_{woc}^{surf}$ from solving Eq.8, we can use Eq.12 to find the corresponding critical number of cycle $N_{crit}$ as a function of $D_g$. For realistic grain sizes (5-100 μm), it is readily seen that $N_{crit} \propto 1/D_g$, in agreement with well known experimental trends, in HCF of austenitic poly-crystals [47].

Eq.12 is plotted in Fig.8 for the same cases (tilt and twist angles) as treated in Fig.7. This representation constitutes a quantitative evaluation of micro-structural barrier effect on the subsequent primary crack propagation (acceleration or retardation), in function of the distance r from the primary crack tip. Such representation can help in both analyzing the simulation results and in making direct simulation/experiment comparisons. For instance, we can compare Eq.12 prediction with micro-crack kinetics as observed in actual poly-crystals. It only requires determining the crystallographic orientations of grain-1 and grain-2 (using electron back-scattering diffraction technique, for example). This ongoing experimental work will be presented separately.

## 5. Conclusions/Summary

Three-dimensional dislocation dynamics simulations are used to study micro-crack interaction with the first micro-structural barrier in fcc bi-crystals loaded in HCF conditions.

It is shown that the primary crack stress field accelerates the surface relief accumulation ahead of the crack tip region (in grain-2), in comparison to that accumulated without a crack. The orientation of the slip systems activated in grain-2 can adopt any combination of tilt and twist angles with respect to the primary crack plane in contrast with simple continuum mechanics descriptions.

If the crack stress field activates exactly the same slip system as found in a "without-crack" simulation, then the cyclic slip further concentrates in the prior substructure, leading to surface relief growth in the initial slip system.

If the active slip system differs between "without-crack" and "with-crack" simulations, prior dislocation substructure has little influence on subsequent extrusion growth.

Grain-1/grain-2 disorientation also determines whether crack tip plasticity is localized or homogeneous ahead of the crack tip. Low-twist (and tilt) orientations are associated with strong localization of the plastic deformation and therefore to faster micro-crack transmission. Large-twist (and tilt) conditions yield broader plastic strain spreading in grain-2 and therefore, slower micro-crack transmission.

Surface evolutions ahead of the primary crack are evaluated using a semi-analytical micro-mechanical model. The model introduces the original idea of the first-barrier compliance. The micro-mechanical model predicts that the number of active slip planes $\mathcal{N}$ increases with the twist angle (for a given $D_g/2r$ ratio). For realistic grain sizes (5-100 μm), the model predicts that micro-crack transmission takes place for $N_{\text{crit}} \propto 1/D_g$, in good agreement with well-known experimental trends. The model in its present form helps in interpreting the simulation results and in making direct simulation/experiment comparisons.

The micro-mechanical model can be improved by developing a quantitative description of the first-barrier compliance in function of the grain-1/grain-2 disorientation and the cyclic loading level. Additional model improvement can also be carried out based on prediction accuracy requirement. For example, 2D crack tip stress field expressions were employed here for simplicity. More accurate expressions for describing a 3D short crack front are available in literature and can be used instead [38]. The micro-mechanical model presented in section 4 is restricted to parallel active planes. Generalisation to multiple slip conditions needs to be undertaken, especially for treating higher cyclic loading amplitudes i.e. low cycle fatigue instead of the high cycle fatigue conditions used here.

## 6. Acknowledgements

The authors acknowledge the financial support of the French National Agency for Research through ANR-AFGRAP project. Dr. Baldev Raj, Former Director IGCAR, Kalpakkam and Dr. S.C. Chetal, Director IGCAR, Kalpakkam are also thanked for their kind encouragements.

**Table and Figure caption**

**Table 1:** Mechanical and microscopic parameters of 316L steel, at T = 300 K.

**Figure 1:** Crack transmission mechanisms between two adjacent surface grains. It is assumed that grain-1 is cracked and that the common grain boundary temporarily blocks further crack propagation. (a) Direct 'bulk' mechanism: micro-crack transmission towards the secondary grain takes place without assistance from the secondary grain surface. (b) Indirect mechanism: micro-crack transmission is assisted by surface displacement accumulation in the secondary grain.

**Figure 2:** Disorientation between crack plane in grain-1 and active slip plane in grain-2. Definition of: (a) tilt angle, (b) twist angle. A third angle, theta, depends on Burgers vector orientation with respect to crack plane in grain-1. Tilt and especially twist angles are found to be far more influential than theta, on the development of grain-2 plasticity ahead of cracked grain-1.

**Figure 3:** Effect of crack tip stress field of grain-1 on active slip system selection in grain-2. (a) Dislocation density evolution of the 2 active slip planes B and C in grain-2 for disorientation angles tilt = -18°, twist = 0°. (b) Normalized resolved shear stress in the different slip systems, with respect to azimuth angle from -90° up to 90°. (c) Active slip plane B and D are normal to the free surface and aligned with azimuth angle where the loading is maximal. Active slip plane C is inclined with respect to the free surface. Dislocations gliding in plane C come across various azimuth angles and therefore, variable amplitude loading, depending on their position in the grain.

**Figure 4:** Effect of crack tip stress field of grain-1 on active slip system selection in grain-2. (a) Dislocation density evolution of the 2 active slip planes D and C in grain-2 for orientation angles tilt = 0°, twist = -35°. (b) Normalized resolved shear stress in the different slip systems,

with respect to azimuth: angles from -90° up to 90° are included in secondary grain-2. (c) Active slip plane of system A6 and D6 is inclined with respect to the free surface. (d) Dislocations gliding in plane D come across various azimuth angles and therefore, variable amplitude loading depending on their position in the grain.

**Figure 5:** Development of crack tip plasticity and associated surface relief in a strained grain ahead of an arrested stage I fatigue crack (10 µm in length). (a) Global cumulated surface displacement $\gamma_{global}$ accumulation with the number of cycles, in cases ❶, ❷ and ❸ (see text for cases descriptions). (b) Typical stress strain curve at the grain scale (global), case ❷ with $\Delta\varepsilon_p = 2\times10^{-4}$. Note the absence of global mean stress and the linear stress accumulation beyond the yield point, characterized by $\Delta\sigma^*_{global,wc}$. (c) Von Mises plastic strain accumulated on the surface after 20 cycles, case ❶, (d) after 36 cycles, case ❷, (e) after 10 cycles, case ❸. Free surface shows cumulated plastic strain along with slip traces; change in colour contrast from green to red near slip traces indicates a higher degree of cumulated strain.

**Figure 6:** (a) Influence of grain disorientation on plastic strain spreading for a given plastic strain range $\Delta\varepsilon_p = 10^{-4}$. a) Tilt = +18°, twist = 0°, plastic strain is localized in only one slip system, in both the surface and in the bulk. (b) Tilt = +35°, twist = 0°, plastic strain is spreads into 2 active slip systems, across the whole grain volume. (c) Tilt = 0°, twist = -35°, d) Tilt = 0°, twist = 90°. Plastic strain spreading sharply increases with the twist angle.

**Figure 7:** Effect of twist angle on the plastic zone size. Dimensionless quantity ($D_g/2r$) is the reciprocal half-distance to the crack tip, plotted with respect to the number of active pile-ups $\mathcal{N}$ in the crack process zone. Twist = 20°, 38°, 71° and 82°. Tilt = 24°±2° in all the cases.

**Figure 8:** Twist angle effect on crack transmission kinetics towards the secondary grain. The curves are normalized with respect to the non-cracked case. A slip ratio $\gamma^{surf}_{wc}/\gamma^{surf}_{woc} > 1$ means extrusion growth acceleration with respect to without-crack conditions; a slip ratio

$\gamma_{wc}^{surf}/\gamma_{woc}^{surf} < 1$ means extrusion growth retardation with respect to without-crack conditions.

The tilt angle = 24°±2° is used in all the cases.